\documentclass[aps, prd, showpacs, twocolumn, nofootinbib, preprintnumbers, floatfix]{revtex4}
\usepackage[utf8]{inputenc}
\usepackage{bm}
\usepackage{latexsym,amssymb,amsmath,amsfonts}
\usepackage{graphicx}
\usepackage{longtable}

\usepackage{bm}

\graphicspath{{fig/}}
\usepackage[usenames]{xcolor}
\usepackage{soul}
\usepackage[breaklinks, backref, colorlinks=true]{hyperref}
\usepackage{tabularx}   
\usepackage{subfigure}
\usepackage{makecell}

\def \tsid {t_\mathrm{s}}
\def \iday {{i_\mathrm{day}}}
\def \sday {{T_\mathrm{s}}}

\begin{document}
 
\title{Very fast stochastic gravitational wave background map-making using folded data}

\author{Anirban~Ain}
\email{ainz@iucaa.in}
\affiliation{Inter-University Centre for Astronomy and Astrophysics (IUCAA), Pune 411007, India}

\author{Jishnu~Suresh}
\email{jishnus@iucaa.in}
\affiliation{Inter-University Centre for Astronomy and Astrophysics (IUCAA), Pune 411007, India}

\author{Sanjit~Mitra}
\email{sanjit@iucaa.in}
\affiliation{Inter-University Centre for Astronomy and Astrophysics (IUCAA), Pune 411007, India}

\begin{abstract}
A stochastic gravitational-wave background (SGWB) is expected from the superposition of a wide variety of independent and unresolved astrophysical and cosmological sources from different stages in the evolution of the Universe. Radiometric techniques are used to make sky maps of anisotropies in the SGWB by cross-correlating data from pairs of detectors. The conventional searches can be made hundreds of times faster through the folding mechanism introduced recently. Here we present a newly developed algorithm to perform the SGWB searches in a highly efficient way. Taking advantage of the compactness of the folded data we replaced the loops in the pipeline with matrix multiplications. We also incorporated well-known HEALPix pixelization tools for further standardization and optimization. Our Python-based implementation of the algorithm is available as an open source package {\tt PyStoch}. Folding and {\tt PyStoch} together has made the radiometer analysis \textit{a few thousand times} faster; it is now possible to make all-sky maps of a stochastic background in just a few minutes on an ordinary laptop. Moreover, {\tt PyStoch} generates a skymap at every frequency bin as an intermediate data product. These techniques have made SGWB searches very convenient and will make computationally challenging analyses like blind all-sky narrowband search feasible. 
\end{abstract}


\maketitle

\section{Introduction}
A new era in astronomy began with the detection of gravitational waves (GW)~\cite{GW150914}. As was anticipated, the first detected sources are all compact binary coalescence~\cite{CBC_O1,GW170104,GW170608,GW170814,GW170817}. GW astronomy, however, promises much more excitement. A vigorous global effort is underway to observe GW signals in widely separated frequency bands. This includes ground-based interferometric detectors aLIGO, aVIRGO, GEO, TAMA, KAGRA~\cite{aLIGO,AdvVirgo,aGEO,TAMA,KAGRA}, pulsar timing arrays ~\cite{IPTA}, and the planned space-based detectors such as LISA, DECIGO, BBO~\cite{eLISA,DECIGO,BBO}. Different kinds of sources~\cite{gw, MTW, ThorneG300} are expected to be seen in the current and future generation network of detectors, the stochastic background being one of the most interesting ones. A stochastic background of gravitational waves (SGWB)~\cite{TanVuk08,CowTan06} can be generated by the superposition of a wide variety of independent and unresolved astrophysical and cosmological sources from different stages in the evolution of the universe. The unresolved compact binary coalescence consisting of black holes and neutron stars, spinning neutron stars, supernovae, cosmic strings, inflationary models, phase transitions, and the pre-Big-Bang scenario are some of these sources which contribute towards this background \cite{AllenSchool, Grishchuk00, Turner96, hotspot, Mazumder2014, SGWB_phase, Cosmic_string}. The background is likely to be anisotropic if it is dominated by the nearby universe~\cite{Mazumder2014, allen97}. The detection of an anisotropic SGWB will offer novel opportunities to study the origin, distribution, and properties of various astrophysical objects, which are still not accessible to the current observational windows of astronomy\cite{AllenSchool,Allen88,sgwb_mbh}.

Various techniques have been developed to search for isotropic and anisotropic SGWB in data from detectors in different bands~\cite{JoeReview}.
For GW detectors, since noise in geographically far away detectors are likely to be uncorrelated, data from pairs of detectors are cross-correlated to look for a stochastic signal~\cite{Michelson87,christ92,flan93,allen97,allen01,lisa_stoch}.
To probe anisotropies of the SGWB using ground-based detectors, the standard technique presently is the GW radiometer algorithm~\cite{LazzariniWeiss,ballmer06,Mitra07}, which is analogous to aperture synthesis often used in radio astronomy~\cite{radio_synthesis,a_synthesis}.
The algorithm has been thoroughly studied and implemented in a pixel~\cite{ballmer06,Mitra07} and spherical harmonic basis~\cite{Thrane09}, where appropriate time-varying phase delays are applied to probe various spatial scales, and is routinely applied to LIGO data~\cite{isoLVC_2009-10,sgwbS5iso,sgwbS4dir,sgwbS5dir,ALLEGRO_limit}. GW can also cause a considerable amount of perturbations in the time of arrival of pulses from millisecond pulsars. Efforts are on to measure these perturbations for some of the well modelled pulsars ~\cite{PTA_multipole, PTA_inj,PTA_sgwb_aniso_results} to constrain stochastic background in the nano-Hertz band, as well as, to probe the anisotropy of the background~\cite{PTA_Joe}.

In a recent work, the efficiency of the GW radiometer algorithm was dramatically improved through the mechanism of data folding~\cite{Ain_Folding}. A temporal symmetry in the GW radiometry algebra was utilized to fold the entire data into one sidereal day, thereby reducing the computational cost by a factor equal to the total number of days of observation. The enormous efficiency brought by folded data allows one to perform new kinds of analyses on long duration signals observed by ground based detectors using much lower computational resources, e.~g., a blind all-sky narrowband search~\cite{eric_ain}. A parallel pipeline developed to implement folding on LIGO data, and the advantages gained from that are well documented~\cite{Ain_Folding}.

However, folding does not exhaust the possibilities for computational improvement of the radiometer analysis. Here we present steps that can boost the efficiency of the algorithm by another factor of few tens. Moreover, the existing implementation of the stochastic pipeline is not equipped to take full advantage of folded data. Since the folded data volume is generally between 1-10 GB, the whole dataset can be loaded into a computer's RAM, thereby dramatically reducing disk access and providing huge I/O advantage.

The scope for improvement is not limited to computational cost and associated conveniences. The current searches for an anisotropic SGWB in LIGO-Virgo data either use a spherical harmonic basis~\cite{sgwbS5dir}, which is appropriate for smooth and diffuse sources, but has limited sensitivity for localized sources like a galaxy cluster, or a pixel space radiometer search on a Cartesian grid. However, analysis of an equal area pixelization scheme is highly desirable for an all-sky map-making application like ours, primarily for better handling of noise in every pixel. Here we present the first implementation of radiometer search that is fully integrated with the Hierarchical Equal Area isoLatitude Pixelization of the sphere (HEALPix) scheme~\cite{HEALPix}, which is perhaps the most widely-used pixelization scheme in astronomy at present for describing all-sky maps and has been rigorously tested by the cosmology community. Apart from the advantages of equal area pixels, HEALPix offers highly efficient tools for Fourier transforms on the sky, making it easy to transform a map from pixel to spherical harmonic bases and vice-versa, thus making the analysis suitable for both localised and diffused sources\footnote{However it is important to note that in practice, the final results (e.g., clean map, upper limit map, etc.) do depend on the chosen basis because of the different kinds of numerical errors associated with the inversion of the Fisher information matrices in different bases.}.

Finally, the existing LIGO-Virgo stochastic pipeline is suitable for broadband searches or targeted narrowband searches. The directional upper limits on GW intensity using data from the first Advanced LIGO observing run (O1) has been calculated~\cite{O1_stoch_direction}. With the fast tools in hand, we wanted to design the implementation in such a way that the intermediate results, the observed maps at every frequency bin, should be derivable in a straightforward way, which can then be combined to get the broadband result. This would also alleviate the need to perform separate searches for different spectral shapes of the modeled power spectral densities of the sources.

This paper is organized as follows. We briefly review the GW radiometer formalism as well as the folding method in Sec.~\ref{section2}. Algorithms used in {\tt PyStoch} and the narrowband map-making process are explained in Sec.~\ref{section3}. Implementation of {\tt PyStoch} and the results are presented in Sec.~\ref{section4}. We conclude the paper with a summary and discussion of algorithm performance in Sec.\ref{section5} along with results from simulated data, whose noise characteristics are statistically identical to that of O1 data.

\section{Mapping the Stochastic Gravitational Wave Background}
\label{section2}

\subsection{Gravitational wave radiometer}

Detectors which are geographically separated by vast distances are expected to have nearly statistically independent noise. Moreover, since a Gaussian stochastic background is characterized by its second moment, weighted cross-correlation of data from two independent detectors turns out to be the optimal statistic for detecting and mapping a SGWB~\cite{allen01,Mitra07}. The GW radiometer algorithm is based on this fundamental principle, which is commonly used to search for an isotropic or anisotropic SGWB in data from the ground-based interferometric detectors.

\begin{figure}[ht]
\centering
\includegraphics{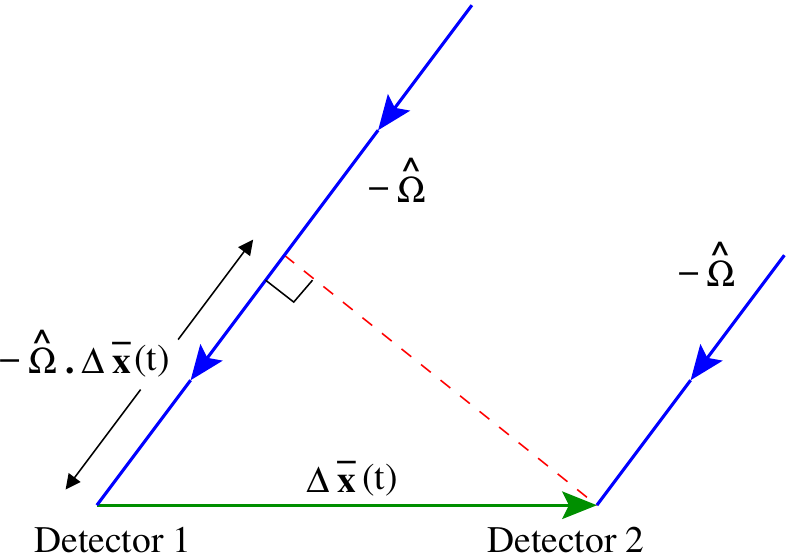}
\caption{Geometry of baseline: schematic diagram of the radiometer. To measure the signal coming from a direction $\mathbf{\widehat{\Omega}}$, data from the detectors are correlated with a phase delay of $\mathbf{\hat \Omega}\cdot {\mathbf{\Delta x}_I (t)}/c$.}
\label{fig_radiometry_1}
\end{figure}
The GW radiometer algorithm considers the delay in the GW signal arrival at detectors which are at different locations. For a given direction in the sky, this delay changes as the baseline orientation changes due to the Earth's rotation. If the time delayed data from two detectors are cross-correlated, the potential GW signals arriving from the given direction interfere constructively while the noises do not. When integrated over a long observational time, the signal cross-correlation grows faster than the noise variance, making the detection statistic more and more significant. The changes mentioned above in baseline orientation due to the Earth's rotation allows one to construct an SGWB map by performing the synthesis for each direction in the sky, with an appropriate choice of the size of the pixel, even with only two detectors. 

The working principle can be explained with the help of Fig ~\ref{fig_radiometry_1}. Consider two gravitational wave detectors denoted by $\mathcal{I}_1$ and $ \mathcal{I}_2$. The output data from the detectors can be written as,
\begin{eqnarray}
s_{\mathcal{I}_{1}}(t) &=& h_{\mathcal{I}_{1}}(t) \ + \ n_{\mathcal{I}_{1}}(t) \label{det_strain1} \, ,\\
s_{\mathcal{I}_{2}}(t) &=&  h_{\mathcal{I}_{2}}(t) \ + \ n_{\mathcal{I}_{2}}(t) \label{det_strain2} \, ,
\end{eqnarray}
where $h_{\mathcal{I}_{1}}(t)$ and $h_{\mathcal{I}_{2}}(t)$ denote the gravitational wave strain in the two detectors due to the SGWB, and $n_{\mathcal{I}_{1}}(t)$ and $n_{\mathcal{I}_{2}}(t)$ denote the noises intrinsic to the first and second detectors respectively. Since the response of a detector (i.e., its antenna pattern with respect to the sky) changes with time, it is convenient to split the data taken from the detectors (time series data) into chunks of duration $\tau$, such that $\tau$ is much greater than the light-travel time between the detectors, but small enough so that the detector response functions do not change considerably over that period. The typically acceptable chunk sizes are a few tens to a few hundreds of seconds. The Fourier transform of such a data segment, often called a \textit{short-term Fourier transforms} (SFTs), is defined as,
\begin{equation}
\label{SFT}
\tilde {s}_I(t;f) \ := \ \int_{t-\tau/2}^{t+\tau/2} \mbox{d}t'  s_I(t')  e^{-i2\pi f t'}  ,
\end{equation}
where a tilde above a variable denotes its Fourier transform of a time series. 
As mentioned before, the noise in different geographically distant detectors can be assumed to be mutually uncorrelated and also uncorrelated with the signal. Thus we have,
\begin{eqnarray}
\langle \tilde{h}^*_{\mathcal{I}_{1,2}}(t;f)  \tilde{n}_{\mathcal{I}_{1,2}}(t;f) \rangle \ = \ 0  ,\\
\langle \tilde{n}^*_{\mathcal{I}_{1}}(t;f)  \tilde{n}_{\mathcal{I}_{2}}(t;f) \rangle \ = \ 0  .
\label{cor_eqns}
\end{eqnarray}
The cross-power spectral density (CSD) of data, $\mathbf{C}^I$, from a baseline $I$ constituted by two detectors $\mathcal{I}_1$ and $\mathcal{I}_2$ and its noise component $\mathbf{n}^I$ are given by, 
\begin{eqnarray}
\mathbf{C}^I \ \equiv \ C^I_{ft} &:=& \widetilde{s}_{\mathcal{I}_1}^*(t;f)  \widetilde{s}_{\mathcal{I}_2}(t;f)\label{eq_CSD} \, ,\\
\mathbf{n}^I \ \equiv \ n^I_{ft} &:=& \widetilde{n}_{\mathcal{I}_1}^*(t;f)  \widetilde{n}_{\mathcal{I}_2}(t;f)\label{eq_noise} \, .
\end{eqnarray}
In the small signal limit, $\langle |\tilde{h}(t;f)|^2 \rangle \ll \langle |\tilde{n}(t;f)|^2 \rangle$, the instantaneous cross-power noise variance becomes,
\begin{equation}
\sigma_{Ift}^{2} \ := \ \langle n^{I*}_{ft} \, n^I_{ft} \rangle \ = \ \frac{\tau^2}{4} P_{\mathcal{I}_1}(t;f) \, P_{\mathcal{I}_2}(t;f) \, ,
\label{eq_cross-power noise_variance}
\end{equation}
where $P_{\mathcal{I}_{1,2}}(t;f)$ are the one-sided power spectral density (PSD) of noise $n_{\mathcal{I}_{1,2}}$ for a segment at time $t$ and $\tau$ is the duration of a segment.

The SGWB is characterized by the CSD of the signal. A search for anisotropic SGWB for a specific spectral distribution $H(f)$ (assuming that the frequency spectral shape is the same in every direction of the sky) boils down to estimation of the SGWB skymap $\mathcal{P}(\mathbf{\widehat{\Omega}})$ that is proportional to the flux coming from different directions on the sky~\cite{Mitra07}. One can however perform the search in different set of basis $e_p(\mathbf{\widehat{\Omega}})$ on the two-sphere, in which the anisotropy map can be expanded as,
\begin{equation}
\label{eq_general basis}
\mathcal{P}(\mathbf{\widehat{\Omega}}) \ := \ \sum_p \mathcal{P}_p  e_p(\mathbf{\widehat{\Omega}})  .
\end{equation}
Then the expectation value of the CSD for a baseline can be written as~\cite{Thrane09}
\begin{equation}
\langle C^I_{ft} \rangle \ := \ \tau  H(f) \sum_p  \mathcal{P}_p  \gamma^{I}_{ft,p} ,
\label{eq_expCSD}
\end{equation}
where $\gamma^{I}_{ft,p} $ is is a geometric factor usually known as the overlap reduction function (ORF) defined as~\cite{christ92,ORF_Finn},
\begin{equation}
\label{eq_overlap_red}
\gamma_{ft,p}^{I} := \sum_{A} \int_{S^2} d \mathbf{\hat \Omega} F^{A}_{\mathcal{I}_1}(\mathbf{\hat \Omega},t) 
F^{A}_{\mathcal{I}_2}(\mathbf{\hat\Omega},t) e^{2\pi i f \frac{\mathbf{\hat \Omega}\cdot {\mathbf{\Delta x}_I (t)}}{c}} e_{p}(\mathbf{\hat \Omega})  ,
\end{equation}
where $\mathbf{\Delta x}_I (t)$ is the separation vector between the two detectors and $F^{A}_{\mathcal{I}_{1,2}}(\mathbf{\hat \Omega},t)$ denotes the antenna pattern functions of the detectors. Thus the observed CSD is a convolution of the SGWB skymap with the kernel $\mathbf{K}$, plus an additive noise term $\mathbf{n}$,
\begin{equation}
\label{eq_CSD_conv}
\mathbf{C}^{I} \ = \ \mathbf{K}^{I} \cdot \bm{\mathcal{P}} \ + \ \mathbf{n}^{I} \, .
\end{equation}
Here $\mathbf{K}^{I}$, the kernel or the beam function is defined as,
\begin{equation}
\label{eq_convolution}
\mathbf{K}^I \ \equiv \ K^{I}_{ft,p} \ := \ \tau \, H(f) \, \gamma^{I}_{ft,p} \, .
\end{equation}
The convolution equation has a standard maximum-likelihood (ML) solution $\mathcal{\hat{P}}_p$ which produces the estimates for the SGWB skymaps~\cite{Mitra07},
\begin{equation}
\label{eq_ML_solution}
\mathcal{\hat{P}}_p \ \equiv \ \hat{\bm{\mathcal{P}}} \ = \ \mathbf{\Gamma}^{-1} \cdot \mathbf{X} \, ,
\end{equation}
where $\mathbf{X}$, the dirty map, is given as,
\begin{equation}
\label{eq_dirtymap}
\mathbf{X} =\frac{4}{\tau} \sum_{Ift} \frac{ H(f) \gamma^{I*}_{ft,p}} {P_{\mathcal{I}_1}(t;f) P_{\mathcal{I}_2}(t;f)} \widetilde{s}_{\mathcal{I}_1}^*(t;f) \widetilde{s}_{\mathcal{I}_2}(t;f) \, ,
\end{equation}
and $\mathbf{\Gamma}$, the Fisher information matrix,  as,
\begin{equation}
\label{eq_fisher}
\mathbf{\Gamma} = 4 \sum_{Ift} \frac{H^2(f)}{P_{\mathcal{I}_1}(t;f) \, P_{\mathcal{I}_2}(t;f)} \,\gamma^{I*}_{ft,p} \, \gamma^{I}_{ft,p'} \, .
\end{equation}
ML estimation of the convolution equation [Eq. (\ref{eq_CSD_conv})]  takes the simple form given in Eq. (\ref{eq_ML_solution}) only when the inverse of the beam matrix exists. So directly inverting the beam matrix is non-trivial. Hence we prefer to solve the linear algebraic equation
\begin{equation}
\label{eq_ML_linear}
\mathbf{\Gamma} \cdot  \hat{\bm{\mathcal{P}}} =  \mathbf{X} \, ,
\end{equation}
to find  $\hat{\bm{\mathcal{P}}}$, the ML estimate for the GWB sky map\footnote{Fisher information matrix $\bm{\Gamma}$ in most cases is not easily invertible. And, due to the noise term in Eq. (\ref{eq_CSD_conv}), the solution of Eq. (\ref{eq_ML_linear}) is noisy. A conjugate gradient or least square method is used to find the best `clean' map $\hat{\bm{\mathcal{P}}}$.}.

\subsection{Folding}
As we mentioned previously, gravitational wave radiometry relies on the basic principles of Earth rotation image synthesis. Characteristic properties of the SGWB and the algebra of analysis technique reveal that there exists a temporal symmetry. In the expressions for the dirty map $\mathbf{X}$ in Eq. (\ref{eq_dirtymap}) and the Fisher information matrix $\mathbf{\Gamma}$ in Eq. (\ref{eq_fisher}), the only two quantities needed for ML estimation of $\bm{\mathcal{P}}$, the geometric part has a period of one sidereal day (i.e. 23 hr 56 min 4 sec). One can use this property to fold the entire detector data of several hundreds of days to only one sidereal day~\cite{Ain_Folding}. This can be done by splitting the time segment $t$ into multiples of sidereal day plus remainder within that sidereal day: $t = \iday \times \sday + \tsid$ and, correspondingly we can rearrange the sum over segments into a sum over sidereal time and a sum over days, $\Sigma_t \rightarrow \Sigma_{\iday} \Sigma_{\tsid}$, where $\iday$ is an integer  representing the sidereal day number in which a given $t$ lies and $\sday$ is the duration of one sidereal day. Then one can rewrite (\ref{eq_dirtymap}) and (\ref{eq_fisher}) as,
\begin{eqnarray}
X_p & = & \sum_{If\tsid} K^{I*}_{f\tsid,p} \, \sum_{\iday} \sigma_{If(\iday\sday + \tsid)}^{-2} \,C^{I}_{f(\iday\sday + \tsid)}  \label{eq_foldx} \, , \ \ \\
\Gamma_{pp'}  & = & \sum_{If\tsid} K^{I*}_{f\tsid,p} \, K^{I}_{f \tsid,p'} \, \sum_{\iday} \sigma_{If(\iday\sday + \tsid)}^{-2}  \label{eq_foldgamma} .
\end{eqnarray}
The summation over $\iday$ part of the above equations correspond to the folded data. It is evident that the folded objects are independent of the search basis indexed with $p$. That is, once the data are folded the same data can be used for fast searches in any basis. 

The algebra gets more involved in practice due to the application of overlapping window functions to reduce spectral leakage. Noise in the neighboring segments no longer remains statistically independent.

A window function is a mathematical function of time, $\mathcal{W}_\mathcal{I}(t)$, that is zero-valued outside of some chosen interval. In GW radiometry, usually a smooth window function is applied to the time series data. Such windows would lead to loss of detector data. In order to prevent that the windows are made to overlap. The usual practice is to use a Hanning window~\cite{O1_stoch_iso} with 50\% overlapping segments. Overlapping windows makes the noise in the neighboring segments partly correlated and the algebra gets more involved. After incorporating the correction in the algebra to account for the effect of overlapping windows, one can express the dirty map and the Fisher information matrix respectively as~\cite{Ain_Folding},
\begin{eqnarray}
X_p &=& \sum_{If\tsid} K^{I*}_{f\tsid,p} \,  x^I_{f\tsid} \, , \label{fwX}\\
\Gamma_{pp'} &=& \sum_{If\tsid} K^{I*}_{f\tsid,p} \big[ K^I_{f\tsid,p'}  \, v^I_{f\tsid} \label{fwGamma} \\
&& - \ K^I_{f(\tsid-1),p'}  \, u^I_{f\tsid} \ - \ K^I_{f(\tsid+1),p'}  \, w^I_{f\tsid} \big] \, . \nonumber
\end{eqnarray}
Here the folded data are given by three real frequency series,
\begin{equation}
\begin{split}
v^I_{f\tsid} \ = &\  \sum_\iday \sigma^{-2}_{If(\iday\sday + \tsid)} \, , \\
u^I_{f\tsid} \ = &\  \sum_\iday \frac{1}{2} \varepsilon^I_{\iday\sday + \tsid-1} \ \times \\
& \qquad \left[ \sigma_{If(\iday\sday + \tsid)}^{-2} \, + \, \sigma_{If(\iday\sday + \tsid-1)}^{-2} \right] \, , \\
w^I_{f\tsid} \ = &\  \sum_\iday \frac{1}{2} \varepsilon^I_{\iday\sday + \tsid+1} \ \times\\
& \qquad \left[ \sigma_{If(\iday\sday + \tsid)}^{-2} \, + \, \sigma_{If(\iday\sday + \tsid+1)}^{-2} \right] \, ,
\end{split}
\end{equation}
\begin{widetext}
and {\em one complex} folded time-frequency map
\begin{equation}
\begin{split}
x^I_{f\tsid} \ = &\ \sum_\iday \left[  \sigma^{-2}_{If(\iday\sday + \tsid)} \, C^{I}_{f(\iday\sday + \tsid)} - \ \frac{1}{2} \varepsilon^I_{\iday\sday + \tsid-1} \, \left\{ \sigma_{If(\iday\sday + \tsid)}^{-2} \, + \, \sigma_{If(\iday\sday + \tsid-1)}^{-2} \right\} \, C^{I}_{f(\iday\sday + \tsid-1)} \right. \\
&- \ \left. \frac{1}{2} \varepsilon^I_{\iday\sday + \tsid+1} \, \left\{\sigma_{If(\iday\sday + \tsid)}^{-2} \, + \, \sigma_{If(\iday\sday + \tsid+1)}^{-2} \right\} \, C^{I}_{f(\iday\sday + \tsid+1)} \right] \, ,
\end{split}
\end{equation}
\end{widetext}
expressed in terms of the redefined CSD,
\begin{equation}
\mathbf{C}^I \ \equiv \ C^I_{ft} \ := \ \frac{1}{\overline{\mathcal{W}_{\mathcal{I}_1}(t) \mathcal{W}_{\mathcal{I}_2}(t)}}  \widetilde{s}_{\mathcal{I}_1}^*(t;f) \, \widetilde{s}_{\mathcal{I}_2}(t;f) \, ,
\end{equation}
and its variance,
\begin{equation}
\sigma^2_{Ift}  \ = \  \frac{\overline{\mathcal{W}_{\mathcal{I}_1}^2(t) \mathcal{W}_{\mathcal{I}_2}^2(t)}}{\left[\overline{\mathcal{W}_{\mathcal{I}_1}(t) \mathcal{W}_{\mathcal{I}_2}(t)}\right]^2} \, \frac{\tau^2}{4} \, P_{\mathcal{I}_1}(t;f) \, P_{\mathcal{I}_2}(t;f) \, .
\end{equation}
The quantity
$\mathcal{W}_I$ a usually a small fraction that depends on the window functions. For 50\% overlapping Hanning windows, which is often the standard choice in the current analyses, $\mathcal{W}_I=3/70$. Here $\widetilde{s}_{\mathcal{I}_{1,2}}(t;f)$ are windowed SFTs and a line over a quantity, e.g., $\overline{\mathcal{W}_{\mathcal{I}_1}(t) \mathcal{W}_{\mathcal{I}_2}(t)}$, denotes average over time~\cite{LazzariniRomano}.
\begin{equation}
\varepsilon^I_{t\pm 1} \ = \ \left\{ \begin{array}{ll}
 \mathcal{W}_I & \mbox{if segment $t\pm 1$ exists for baseline $I$}\\
 0 & \mbox{otherwise}
 \end{array} \right. .\nonumber
\end{equation}

Using the deconvolution technique discussed previously, we will then produce the clean map out of the dirty map with the help of Fisher matrix. The science is carried out by producing the clean skymaps. In ~\cite{Ain_Folding} we have discussed in detail about the advantages of the folding algorithm. The advantages include \textit{efficiency}: computational resources reduces by a factor of few 100; \textit{portability}: folded data size is $\sim 1.3$GB for standard stochastic search; \textit{management}: irrespective of the observation time computation time is fixed; \textit{modularity}: one can do intensive folding part in a low-level language like C and complex algebra of filtering for different searches in MATLAB or Python; and \textit{convenience}: possible to analyse in a portable computer. In this scenario, we are aiming towards developing a new pipeline, which is capable of using the folded data to produce the skymap in pixel basis with a significant improvements. 

\section{Efficient Mapmaking Algorithm: {{\tt P\lowercase{y}S\lowercase{toch}}}}
\label{section3}

A variety of data analysis techniques have been proposed and implemented in the past for the SGWB searches. In the recent times, the searches for anisotropic backgrounds are being performed either in spherical harmonic basis, by measuring the first few spherical harmonic multipoles of the sky, or in the pixel basis with an equispaced grid in latitude and longitude (i.e., the pixel are not of equal area). The {\tt PyStoch} pipeline is an attempt to develop a new implementation incorporating the Hierarchical Equal Area isoLatitude Pixelization (HEALPix) scheme~\cite{HEALPix}, which makes it possible to trivially obtain the spherical harmonic moments as well. {\tt PyStoch} also generates maps at every frequency bin as intermediate results. {\tt PyStoch} uses some existing PyCBC modules~\cite{pycbc_1,pycbc_2} and is designed to take full advantage of folded data, which, in addition to the speed-up resulted from folding, leading to nearly a factor of hundred boosts in the computational efficiency.

\subsection{Narrowband maps}
From Eq.~(\ref{eq_CSD}), Eq.~(\ref{eq_cross-power noise_variance}) and Eq.~(\ref{eq_dirtymap}), it is evident that one can split the expression for dirty map into a frequency sum and a time dependent sum,
\begin{equation}
X_p \ = \ \tau  \sum_{f} H(f) \sum_{It}  \gamma^{I*}_{ft,p} \sigma^ {-2}_{Ift} C^I_{ft} , 
\end{equation}
such that, the broadband map $X_p$ becomes a source spectrum weighted sum of narrowband maps $X_{p,f}$,
\begin{equation}
X_p \ = \ \sum_{f} H(f) \, X_{p,f} \, ,
\end{equation}
where the expression of the narrowband maps at each frequency is given by,
\begin{equation}
X_{p,f}\ = \ \tau   \sum_{It}  \gamma^{I*}_{ft,p} \sigma^ {-2}_{Ift} C^I_{ft} . 
\label{eq:narrowband}
\end{equation}
In the existing pipeline, the summation over time follows the summation over frequency because the length of data is arbitrary. This order is non-trivial to change once the pipeline is fully developed. {\tt PyStoch} is designed to do the time summation first and hence the narrowband maps are automatically produced as an intermediate product.
Computation of the Fisher information matrix for each frequency bin can also be split in this manner, indicating that it will be possible to deconvolve the narrowband maps or to produce upper limit maps at each frequency saving the extra computation power that was necessary for making these maps.

\subsection{HEALPix}

HEALPix is arguably the most popular equal area pixelization scheme in modern Astronomy. Equal area pixelization makes it easier to track pixel noise and its covariances. HEALPix has been primarily developed by the Cosmic Microwave Background community for two decades\cite{HEALPix}. In HEALPix, the 2-sphere is tessellated into $12 n^{2}_{\mbox{side}}$ pixels where $n_{\mbox{side}}$ is an integer power of $2$ which defines the number of divisions along the side of a base-resolution pixel that is needed to reach a desired high-resolution partition. By considering the optimal resolution required for the radiometer analysis for the two LIGO detectors in the US, we choose  $n_{\mbox{side}}=16$ which corresponds to a pixel width of $\sim 3^{\circ}$ and 3072 pixels for the full sky. The advantage of using pixel basis is that one need not worry about the loss of information as compared with the analysis in spherical harmonics basis.

\begin{figure}[ht]
\centering
\includegraphics[width=0.5\textwidth]{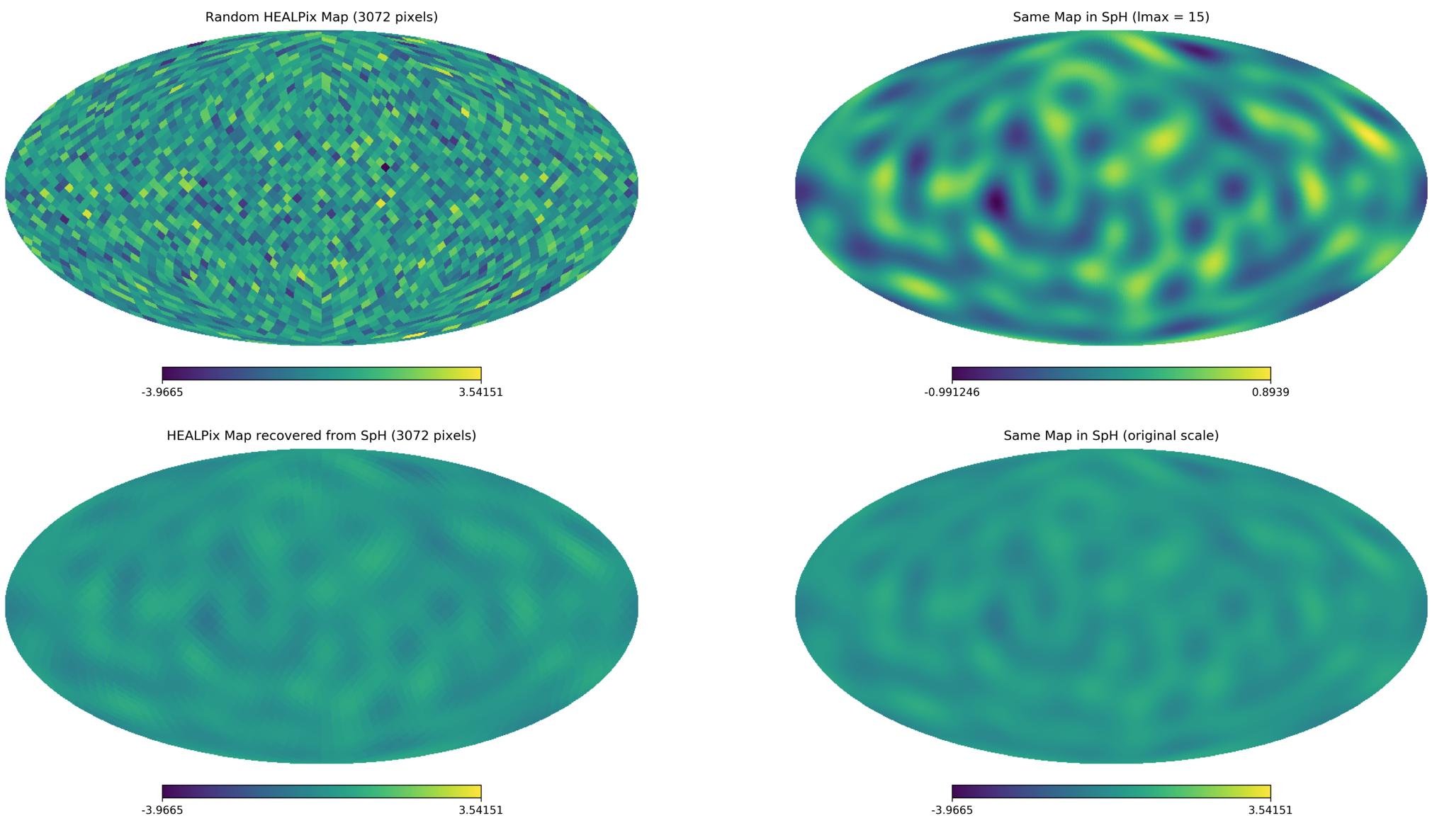}
\caption{Illustration of the fact that HEALPix is more efficient in detecting localized sources. The top left image is a random HEALPix map ($\mbox{n\_side}=16$), the top right is the same map in spherical harmonics ($\mbox{n\_side}=16$). The bottom two maps are produced from the top right map but the `colour scale' is same as the top left map. The bottom left map is in HEALPix (converted from the top right) the bottom right one is same as the top right but with different colour scale. It is evident that spherical harmonics cannot retain the localized pixel information.}
\label{pixsph}
\end{figure}

Figure~\ref{pixsph} shows how a spherical harmonics basis suppresses pixels with extreme values. The top left image has more information than the spherical harmonic in the bottom right. This is obvious because the HEALPix map ($\mbox{n\_side}=16$) has 3072 pixels with unique information in each pixel, but the spherical harmonics up to $l_{\mbox{max}}=15$ has only 256 components.

Furthermore, HEALPix offers ready-made tools to provide spherical harmonic transform of a pixel-space map and vice-versa, via Fast Fourier Transform (FFT) by taking advantage of isoLatitude pixelization. Hence a separate search for anisotropic SGWB in a spherical harmonic domain may become redundant. We have used here the {\tt healpy} package, the Python implementation of HEALPix.

\subsection{Posing mapmaking as a matrix multiplication}

Taking advantage of the compression achieved by folding, {\tt PyStoch} is explicitly calculating $ K^{I*}_{f\tsid,p},  x^I_{f\tsid}$ for all time segments and performing Eqs.~(\ref{fwX},\ref{fwGamma}) as a matrix multiplication for a fixed frequency $f$. This provides much more efficiency compared to the usual practice of looping over time segments and computing each component in place. When using folded data for the sky map making, the segment times and data lengths are predictable. Every data-segment can have previously determined start times, as the segment division is now in sidereal times. This allows pre-calculation of the overlap reduction function. Hence it reduces computational time significantly compared to the usual method of calculating the overlap reduction function (ORF) on-the-go for all available segments. This approach makes it easy to perform the analysis for a network of detectors by calculating the dirty map and Fisher matrix ($X_{p},~\Gamma_{pp'}$) for all detector pairs in the network and adding them.

\subsection{Efficient computation of overlap reduction function}

Calculation and storage of ORF for a sidereal day with a segment duration of $52$ sec and a bin size of $0.25$~Hz requires almost $292$~GB of RAM for a resolution corresponding to $n_{\mbox{side}} = 16$. {\tt PyStoch} alleviates this high RAM requirement issue by introducing seed matrices for the ORF, which can be used to compute the ORF in a fast manner, alleviating the demand for an unusually large amount of memory.

\begin{figure}[ht]
\centering
\includegraphics[width=0.5\textwidth]{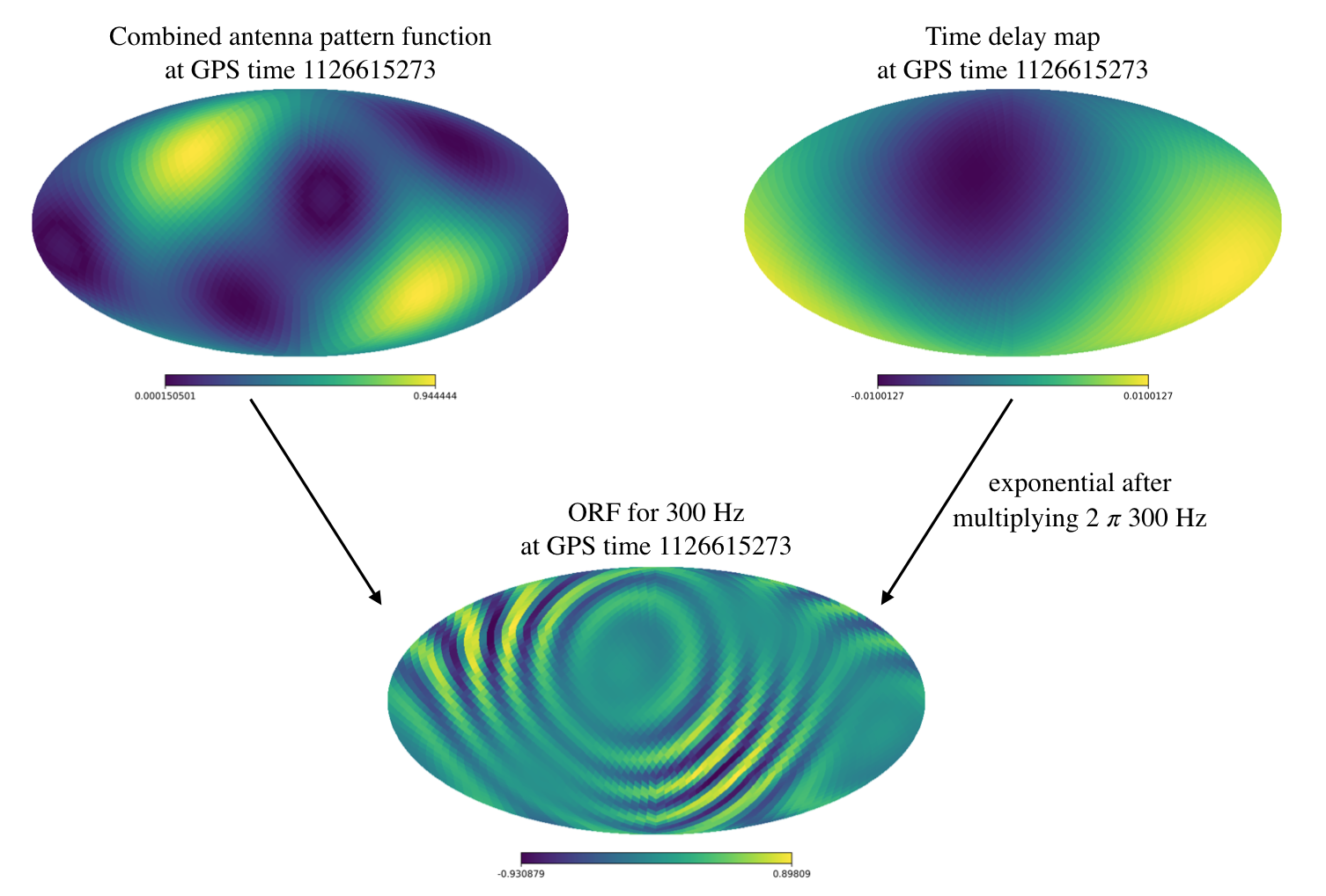}
\caption{The top left map is essentially the sensitivity map for a baseline for a particular time. The top right map is the time delay map for the same time. Multiplying the top right map with a frequency dependent term ($2 \pi i f$) and taking the exponential then multiplying it pixel by pixel with the top left map we get the ORF for that time and that frequency. Here only the real part of the ORF is mapped in the bottom map. We call the top two maps ORF seeds. The ORF seeds for all available time segments are called ORF seed matrices (each row of the ORF seed matrices corresponds to ORF seeds for a particular segment).}
\label{orf from seed}
\end{figure}

Let us consider the algebraic structure of ORF. It can be seen in Eq.~(\ref{eq_overlap_red}) that the ORF depends on two components, the combined antenna pattern function, $F^{A}_{\mathcal{I}_1}(\mathbf{\hat \Omega},t)  F^{A}_{\mathcal{I}_2}(\mathbf{\hat\Omega},t)$, and the time delay, $\mathbf{\hat \Omega}\cdot {\mathbf{\Delta x}_I (t)}/c$, which have no frequency dependence and they are sky maps for a particular time segment. We call these two quantities the ORF seeds, which are easy to store in the RAM and can be used to calculate the ORF for each time segment and frequency bin. 
The combined antenna pattern function provides all possible combination of antenna pattern function corresponding to different detectors (in this demonstration we have used two), whereas from the phase factor expressed in the overlap reduction function one can find the time delay corresponding to each pixel. The quantity ${\mathbf{\hat \Omega}\cdot {\mathbf{\Delta x}_I (t)}}/{c}$ gives the time delay between two detectors in receiving a signal from a certain direction $\hat \Omega$. The process is pictorially shown in Fig.~\ref{orf from seed}. 

%

\subsection{Putting the pieces together}

In this pipeline, our first step is to go through the folded data, produced by folding the entire data of LIGO's observational run to a single sidereal day, and produce the combined antenna pattern function map for every time segment in the folded data. One can store this combined antenna pattern function maps as a matrix which has a dimension equal to the number of pixels times the number of time segments. Similarly, for the time delay maps one can calculate and store the delay between two detectors for different baselines as maps.  This can be considered as calculating and storing the time delay for different time segments in a matrix having a dimension of the number of time segments times number of pixels. One can calculate and store these seed maps within no time as matrices using a small amount of RAM. This entire process of calculating and storing these quantities took only  $155$~MB of RAM each.  As a result, the $3314$~pairs of seed maps corresponding to one sidereal day's data with $52$~sec segment duration could be produced on a laptop in just $20$~seconds.

From these pairs of seed maps, by considering both the combined antenna pattern function map as well as the exponential of the time delay map, for each frequency one can calculate the overlap reduction function. This can also be stored as a matrix with dimension equal to the number of frequency segments times number of time segments.
 
Once we have the ORF, calculation of the dirty map is straightforward. The CSD $C^I_{ft}$ and PSD $\sigma_{Ift}$ are stored in two different channels in stochastic intermediate data (SID) or folded SID (FSID). The CSD and PSD are in the frequency domain, where the minimum and maximum frequencies and the frequency resolution of CSD and PSD determine how many frequency bins are there. We have to loop over all the frequency bins, and each iteration of the loop will produce one narrowband map for that frequency.

In each loop, the ORF is calculated from the ORF seeds for the particular frequency corresponding to that loop. Then the columns in CSD and PSD for that frequency are multiplied with the ORF. As per Eq.~(\ref{eq:narrowband}), the result is the narrowband map for that frequency. Looping over all the frequencies gives all the narrowband maps.

\section{Implementation and Results }
\label{section4}
We have implemented the {\tt PyStoch} code on LIGO's first observational run (O1) data from the Hanford and Livingston detectors We have also used it for simulated data.
Raw data from the detectors are down-sampled to 4 kHz and cross-correlated. Cross-correlation is done in the frequency domain for faster calculation. 
We then used the folding code to fold the data into one sidereal day. The folding code can take care of the overlap of the segments and data quality cuts, but the data we used already had it done during the cross-correlation. The output of folding is $3314$~frames of $52$~sec segment duration which span a complete sidereal day. 

{\tt PyStoch} reads the parameters which are the same for all the frames (e.g., GPS start time, frequency cutoffs, segment durations, etc.) from the first frame. Then it reads the CSD and PSD from all other frames. We had our data in a spectrum where the lower and higher cutoff was 20 Hz and 1980 Hz respectively, and it had a resolution of 0.25 Hz.
The HEALPix map resolution we choose corresponds to $n_{\mbox{side}} = 16$, which uses 3072 pixels for the entire sky (each pixel covers approximately 13 square degrees of the sky). The overlap reduction function had to be calculated for all the $3314$ time segments. Each of the two sets of seed matrices was hence a matrix of dimension $3072\times3314$ with each element a real number. The memory consumption of the seed matrices is 156MB. On a typical laptop (2.6 GHz processor, 4 CPU threads) it takes less than 20 seconds to calculate and save the seed matrices.

\begin{figure}[ht]
\centering
\includegraphics[width=0.225\textwidth]{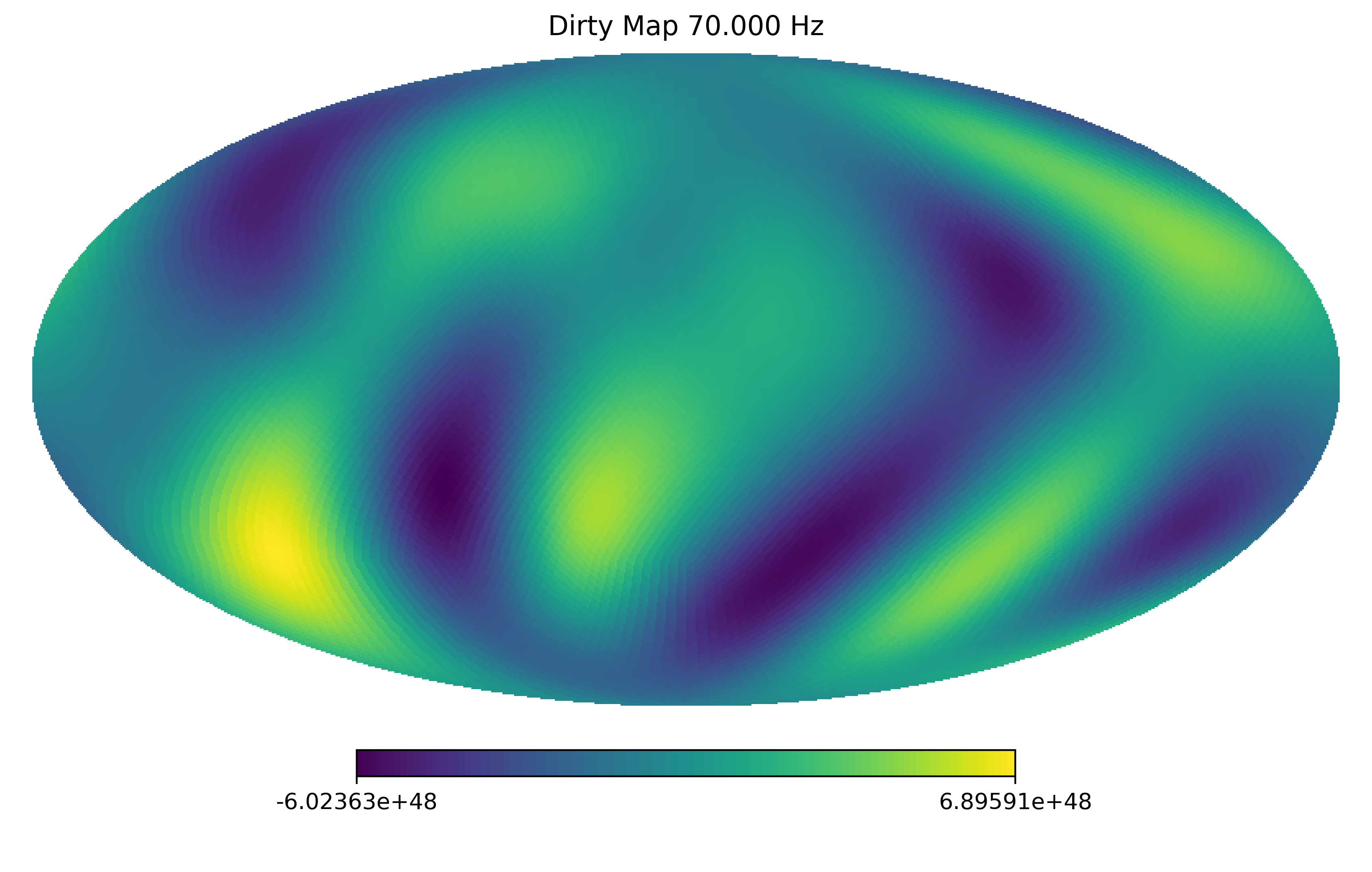}
\includegraphics[width=0.225\textwidth]{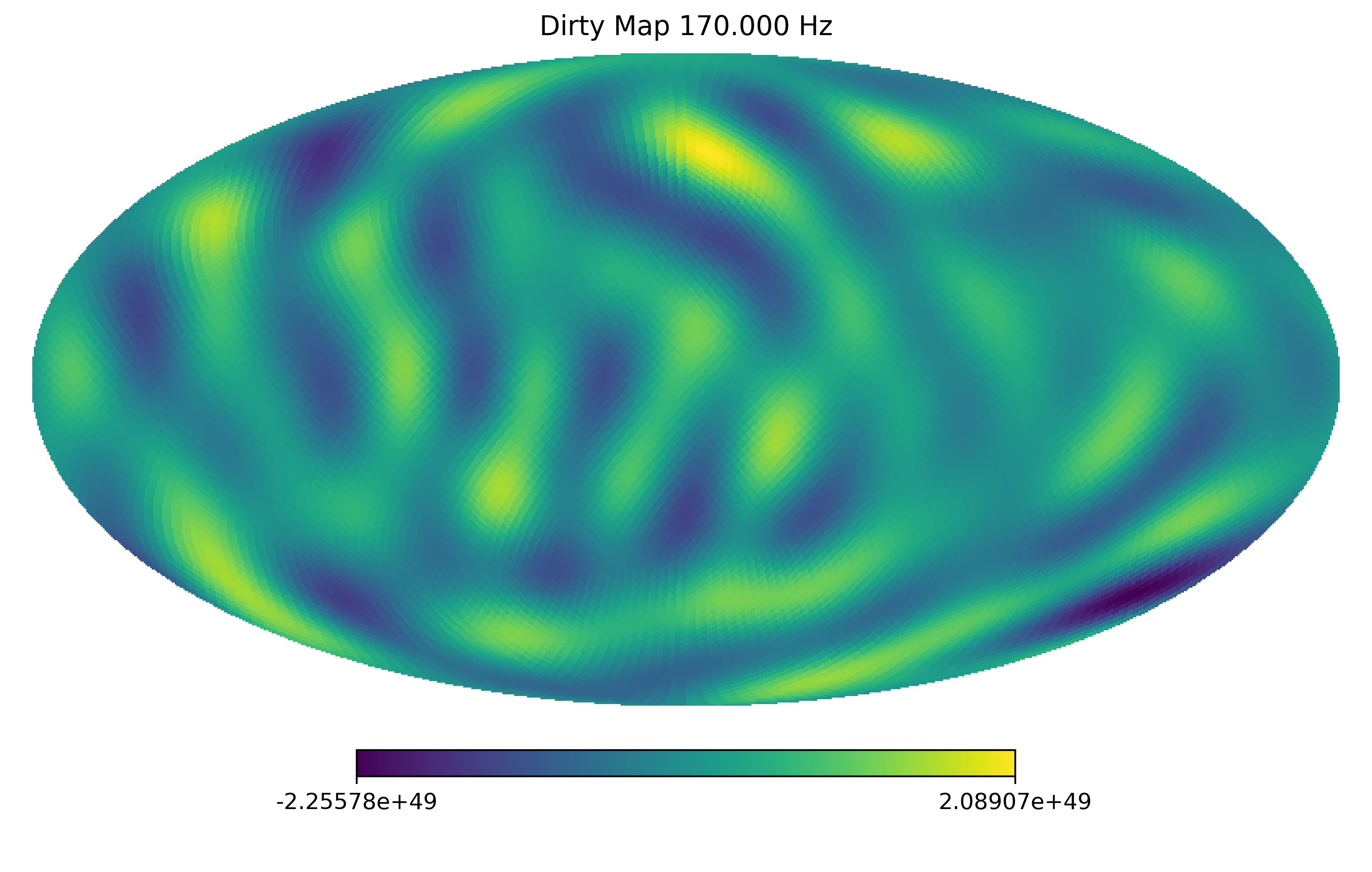}
\includegraphics[width=0.225\textwidth]{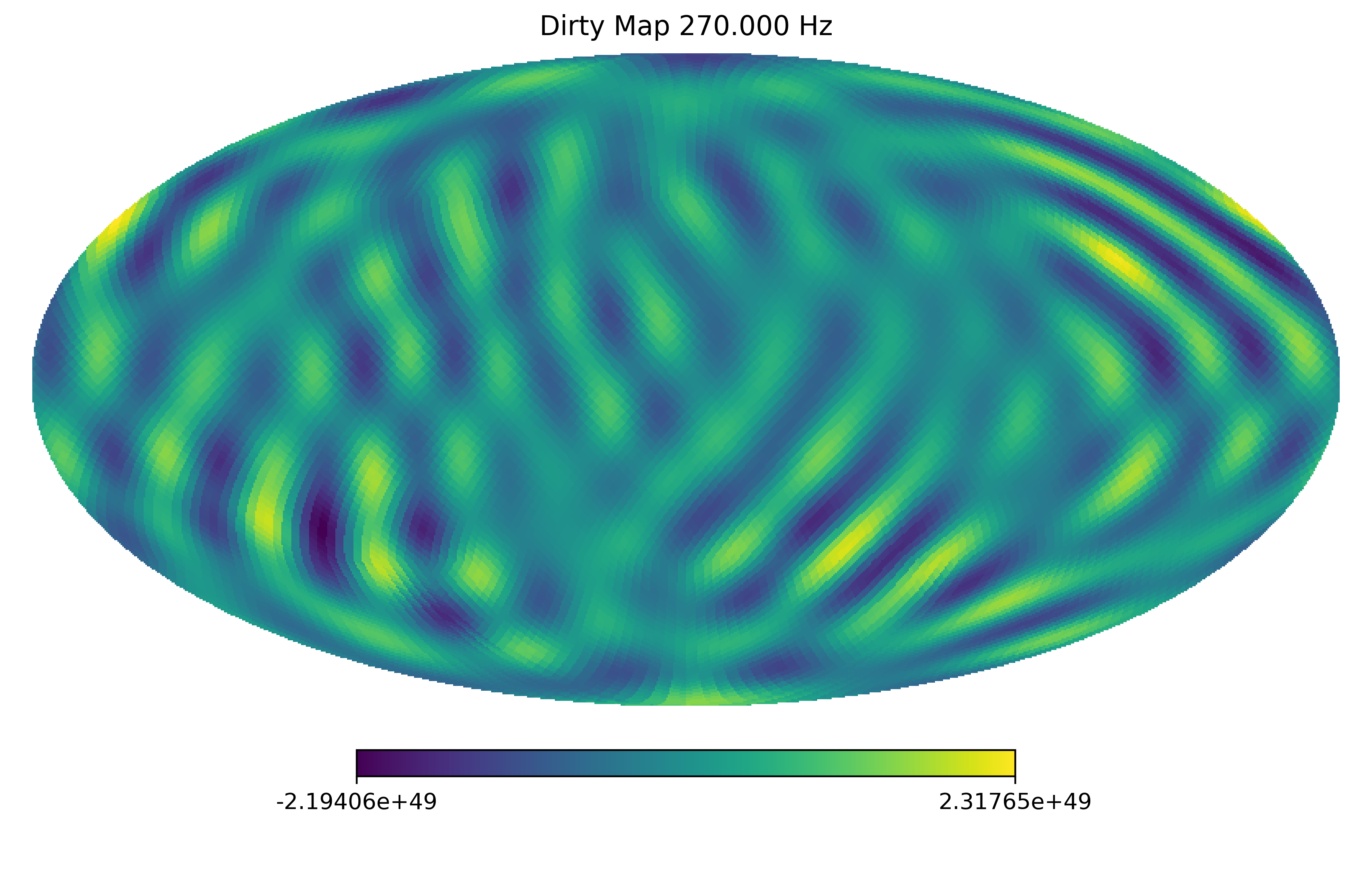}
\includegraphics[width=0.225\textwidth]{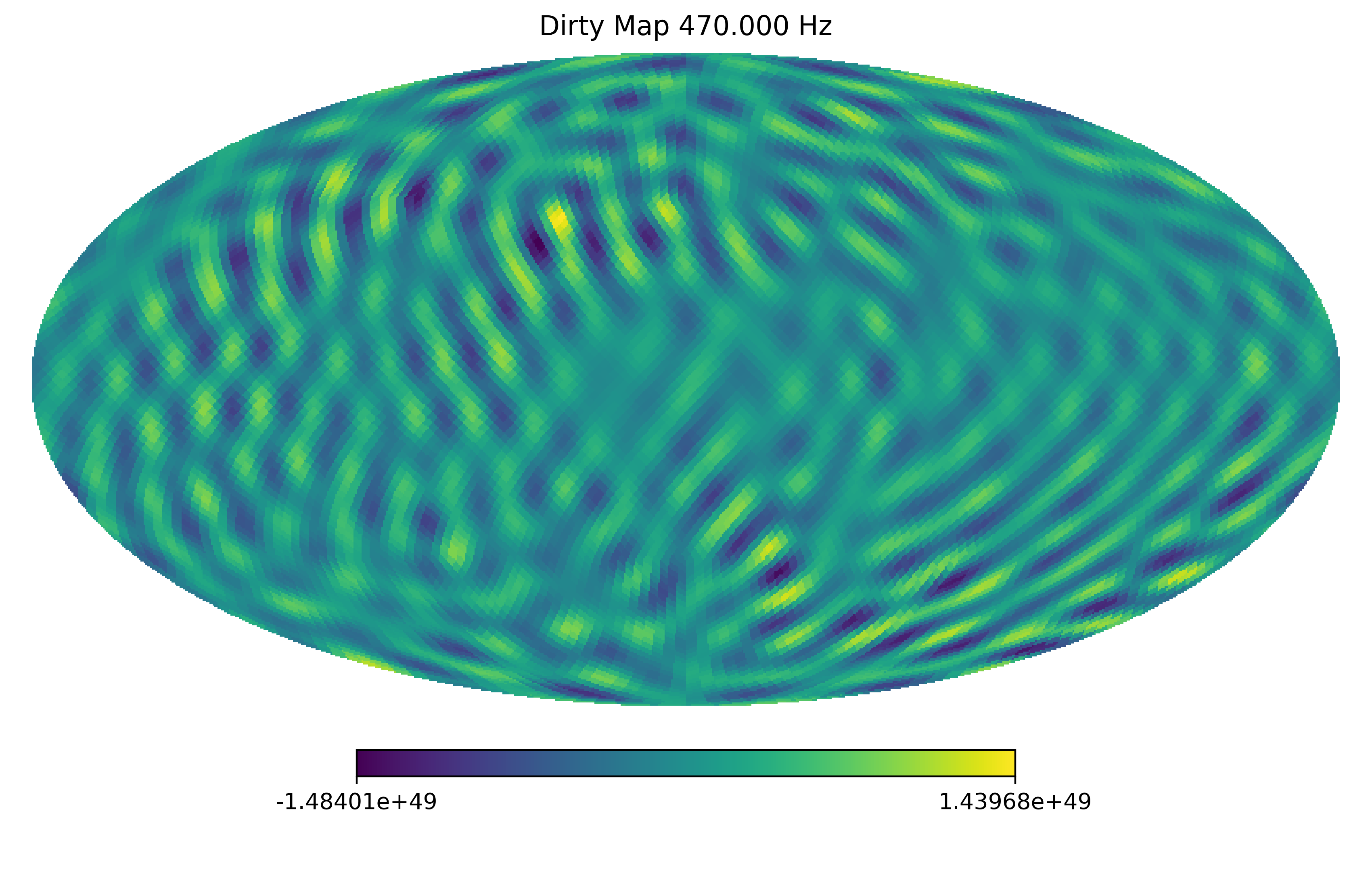}
\caption{\label{fig:narrowband_maps}Sample narrowband dirty maps from simulated data at four different frequencies 70, 170, 270 and 470 Hz are shown. The simulated data has the same statistical properties as O1 data. In these plots it can be seen that the spot sizes get smaller as the frequency goes up due to the diffraction limit. }
\end{figure}
While calculating the narrowband maps, we restricted ourselves to an upper cutoff of 500 Hz, so we had 1920 frequency bins. Then we load the CSD ($C_{Ift}$) and PSD ($\sigma_{Ift}$). The appropriate $1920$ frequency bins are taken for the calculations so the data (CSD and PSD) are matrices of $1920\times3314$.
We now have to loop over $1920$ frequencies. In each loop for a particular frequency, ORF is calculated using the ORF seed matrices. Then we take one column from the CSD and PSD matrices which is of the size $1\times3314$. This corresponds to the data for that frequency and all time segments. When this data column is multiplied with the ORF, we get a $3072\times1$ matrix. This matrix is the narrowband map for that frequency. In Fig.~\ref{fig:narrowband_maps}, $4$ of the $1920$ narrowband maps generated from simulated data are shown. This simulated data which we used is generated by adding a random phase factor to the O1 CSD for all frequency bins. This preserves the statistical properties of the data but renders the results unphysical.

\begin{figure}[ht]
\centering
\includegraphics[width=0.225\textwidth]{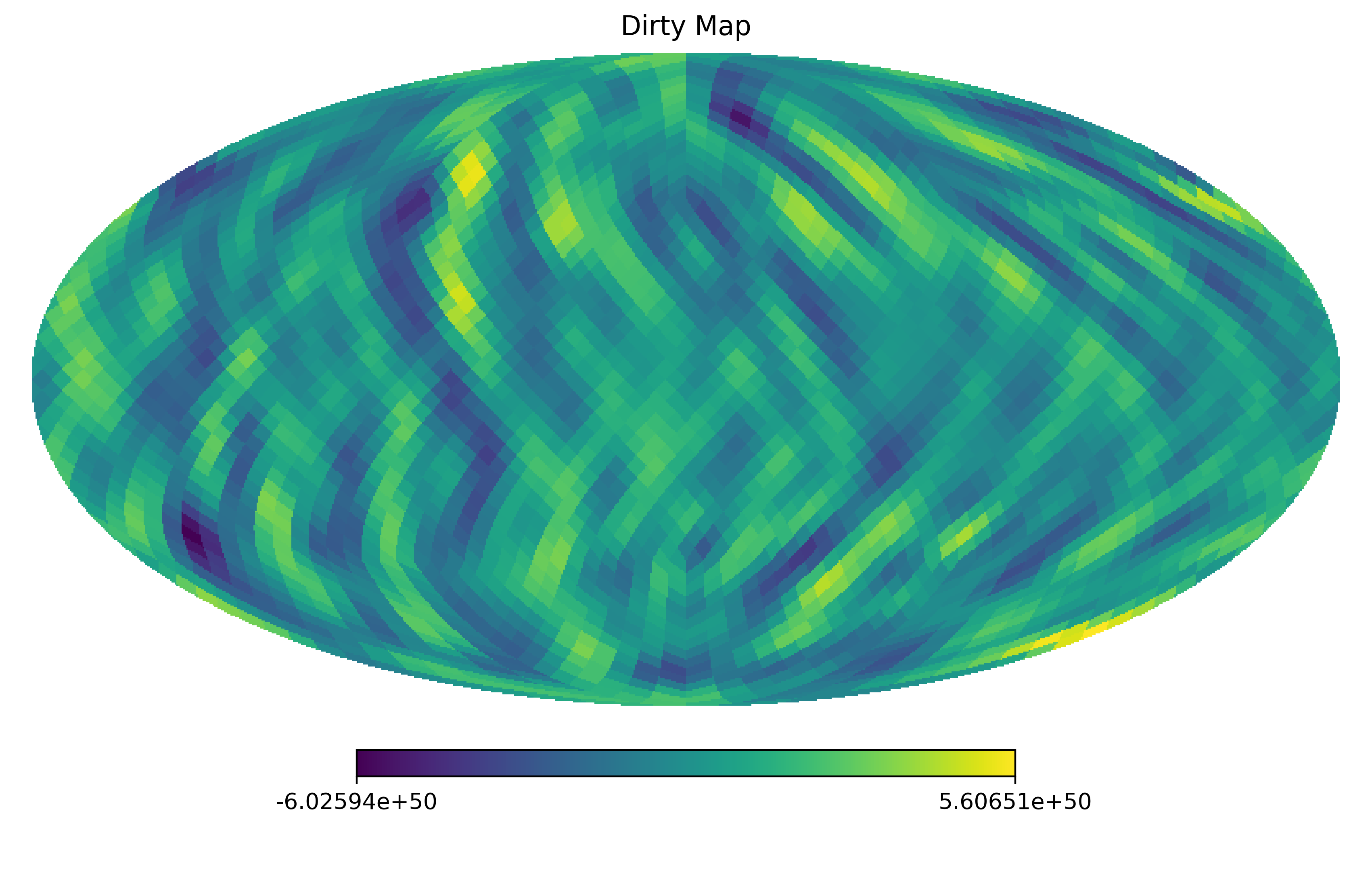}
\includegraphics[width=0.225\textwidth]{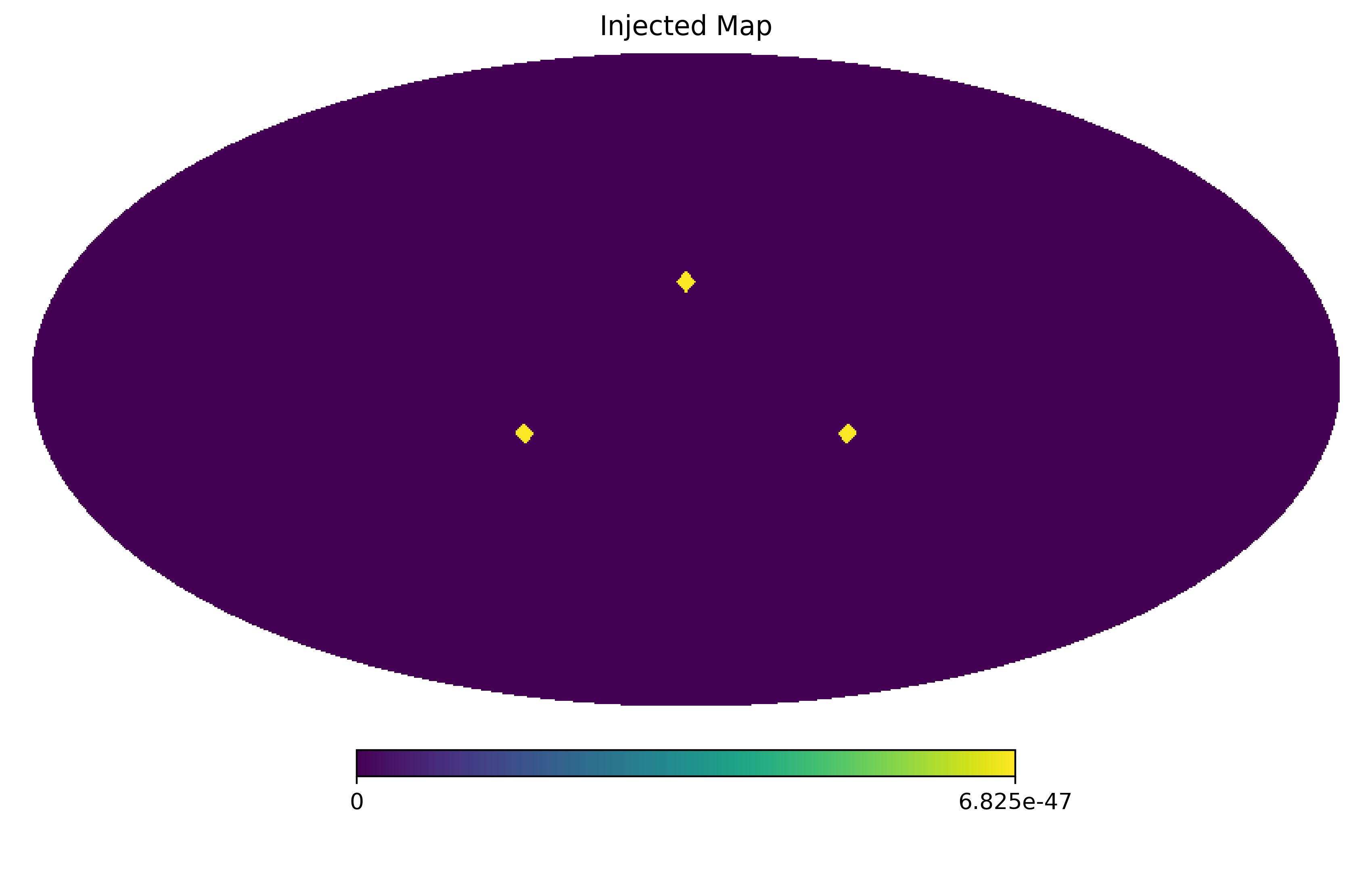}
\includegraphics[width=0.225\textwidth]{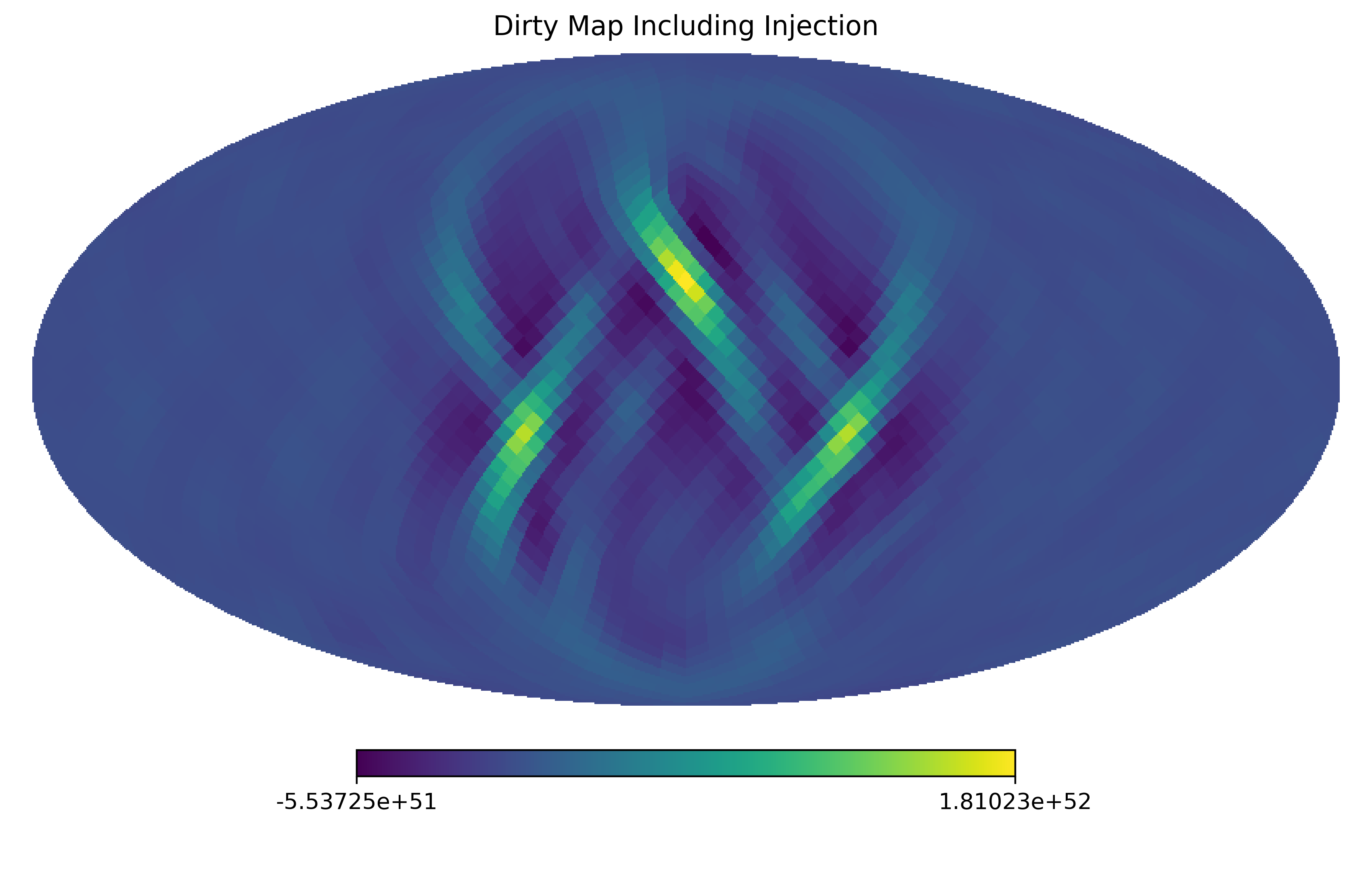}
\includegraphics[width=0.225\textwidth]{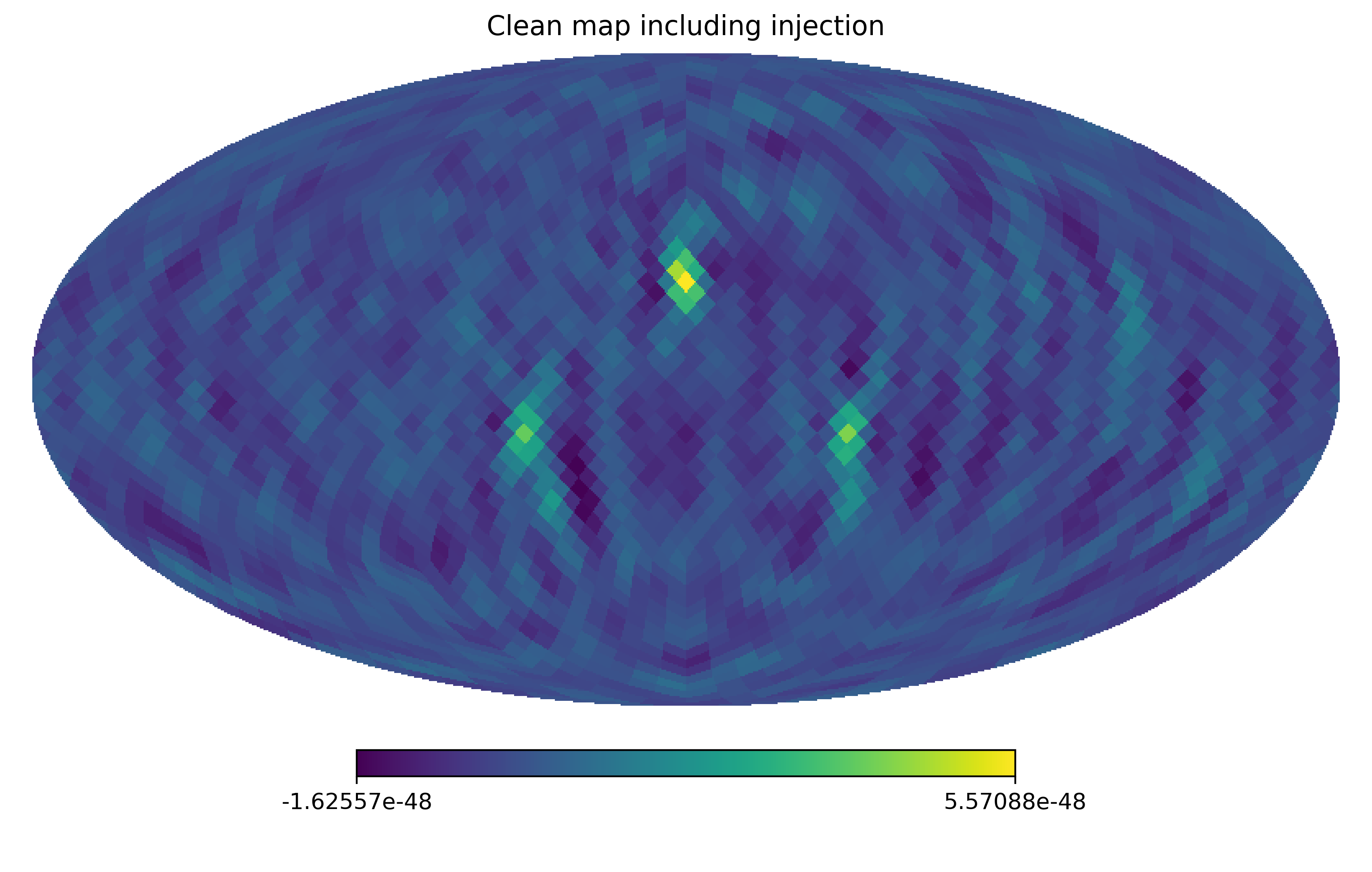}\caption{\label{fig:point_source_inj}The top left map is broadband dirty map from simulated data, the top right map shows the injected sources. The bottom left map is the dirty map made from simulated data including the injections. The bottom right map is the clean map obtained by deconvolution of the (bottom left) dirty map including injections.}
\end{figure}
To further validate {\tt PyStoch}, we tested the code with some injections. The injection sources we used were broadband point sources with a flat spectrum at three different points of the sky, the strengths were two orders higher than the noise PSD. We calculated the CSD for these injected sources and added them to the CSD data. After running the code, the injected sources were properly recovered. The results from applying this injection on simulated data are shown in Fig.~\ref{fig:point_source_inj}. This figure also includes a map in the bottom right which is obtained by deconvolution of the dirty map with a Fisher matrix. This was a check to see if the source strength and location in the clean map match the injected map. In this case, the point sources in the clean map appears at the same location as the injected map, thus validating the deconvolution method. Moreover, the point sources in the clean map are more localized than in the dirty map.

\section{Conclusions}               
\label{section5}
The primary advantage of {\tt PyStoch} is the speed up and convenience. It makes the map calculation few hundred times faster\footnote{Tens of times faster on a single thread, hundred times faster when used with multi-threading.}. Table~\ref{PyStoch_adv} shows the scale of speed up by folding and {\tt PyStoch}. It also gets rid of the requirement for storage to save intermediate results. With folding and {\tt PyStoch} SGWB searches with LIGO data can be done on a laptop, in place of parallel computing on few hundred processors, which is very convenient.

\begin{table}[ht]
\begin{tabular}{|m{0.1\textwidth}|c|c|c|}\hline\hline                                      &\thead{Conventional\\Pipeline} & \thead{Folding\\Pipeline} & \thead{Folding and\\{\tt PyStoch}} \\ \hline
Intermediate Data   & 450 GB   & 1.5 GB & 1.5 GB   \\\hline
Processing Time      & 10 CPU years          & 10 CPU days      & 40 CPU minutes      \\ \hline
Intermediate Results & 800 TB                & 2.5 TB           & not required        \\ \hline
Final Results    & 500 MB                & 500 MB           & 500 MB              \\ \hline\hline
\end{tabular}
\caption{This table shows the estimated calculation time (on a single node of IUCAA computational facility) and storage required to calculate the narrowband maps using three pipelines: the standard pipeline, the standard pipeline with folded data and {\tt PyStoch} with folded data.}\label{PyStoch_adv}
\end{table}

Another advantage is that {\tt PyStoch} produces results regarding narrowband maps. In the older pipeline one had to specify the expected SGWB spectrum $H(f)$ before running the pipeline. But {\tt PyStoch} does not require the spectrum. From Eq.~(\ref{eq:narrowband}), the set of narrowband maps, the spectrum-specific result can be produced using the following equation,
\begin{equation}
X_p  =  \sum_{If} H(f) X_{f,p}
\end{equation}
The above summation of narrowband maps can be done in one matrix multiplication. Also, {\tt PyStoch} is a directed search for all directions in the sky, which means that if one wants to search a particular direction of the sky, e.g., Sco X-1 or Virgo cluster, it is straightforward. We only have to see which pixel(s) include the source and we can just add the pixels (since HEALPix pixels corresponds to the equal area in the sky one does not even have to worry about pixel weights). Similarly, {\tt PyStoch} search results can be contracted into the result of isotropic search just by adding all the pixel values. All data quality cuts, removing bad frequencies (notch list) and correction for the windowing and the segment overlaps can be incorporated during or before running {\tt PyStoch}.

\begin{figure}[ht]
\centering
\includegraphics[width=0.5\textwidth]{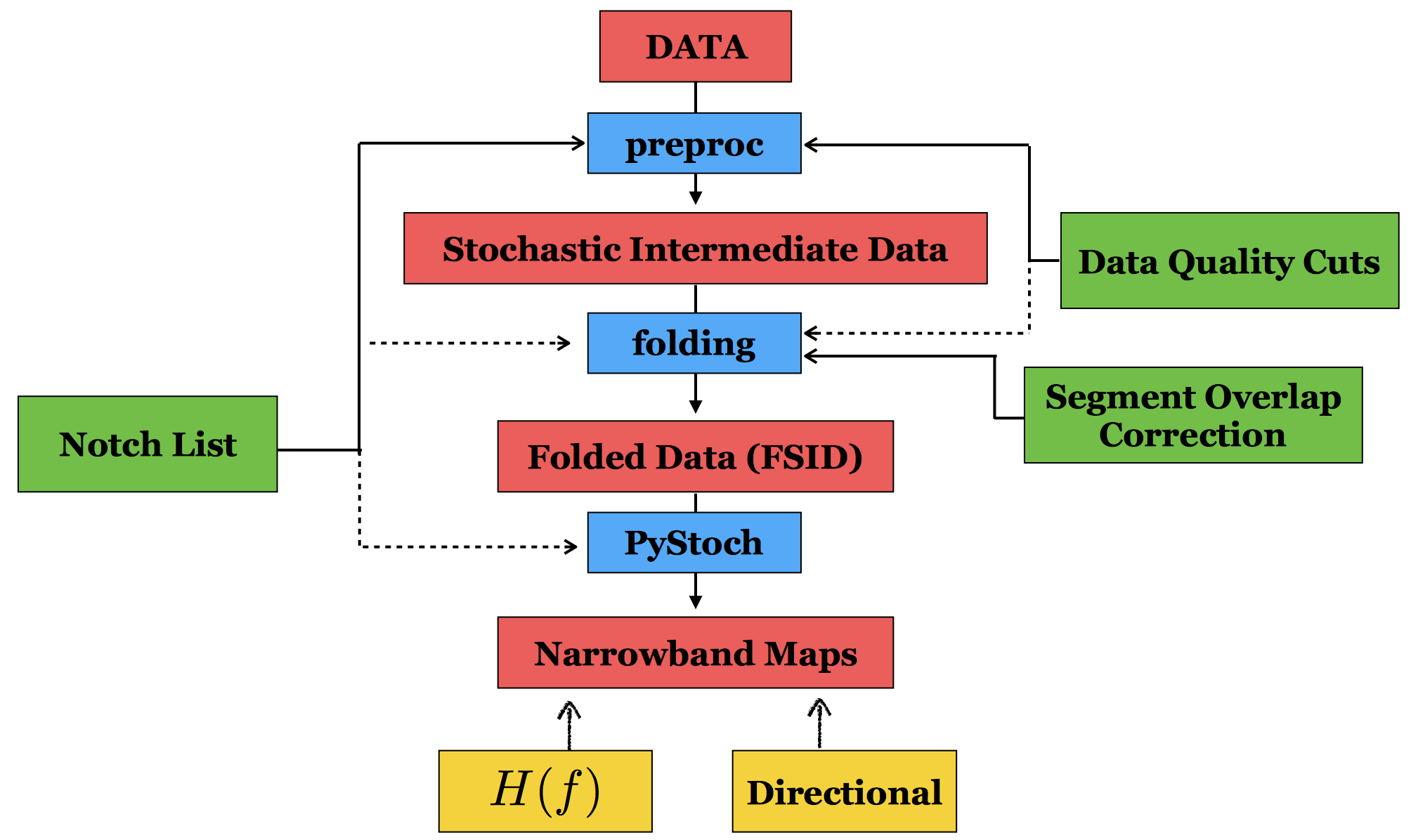}
\caption{\label{fig:PyStoch_algo} Flowchart for a pipeline including folding and {\tt PyStoch}. The interferometer data are cross-correlated in the first step preproc (short for pre-processing), turning the data into SID. The SID is then folded into FSID by the folding module. {\tt PyStoch} takes the FSID and calculates the narrowband maps. Directional or spectrum specific searches can be performed on those narrowband maps. Corrections for bad frequencies (notch list), segment overlaps, data qualities can be applied at many points in this pipeline (solid arrows indicate where we applied it, dotted arrows indicate other modules where it can be applied).}
\end{figure}

The enormous efficiency achieved by folding is further enormously amplified by {\tt PyStoch}. Folding makes the analysis few hundred times faster, {\tt PyStoch} also makes the analysis few ten times faster. So our new algorithm for stochastic analysis (Fig.~\ref{fig:PyStoch_algo}) is about {\em ten thousand} times faster than the standard pipeline, and the results are more general and versatile.  {\tt PyStoch} automatically uses multi-threading when utilized by a multi-core CPU. This can make the SGWB search even faster. Also, this new pipeline eliminate the need for intermediate data storage. The results of {\tt PyStoch} are more practical. In future, {\tt PyStoch} will enable more in-depth searches and maps of SGWB.

{\tt PyStoch} uses PyCBC routines to calculate the ORF. PyCBC has information about other detectors (aVIRGO, GEO600, KAGRA, etc.~\cite{AdvVirgo, GEO600, KAGRA}) incorporated into it. This makes it very easy to tweak {\tt PyStoch} for baselines other than the Livingston-Hanford baselines. We have tested it for many different baselines. As a toy model, let us consider the situation where VIRGO becomes as sensitive as AdvLIGO and when LIGO-India starts operating. In these scenarios, {\tt PyStoch} can be easily used to map the SGWB using the global network of terrestrial detectors.
\section{Acknowledgment}

We thank the LIGO Scientific Collaboration for access to the data and gratefully acknowledge the support of the United States National Science Foundation (NSF) for the construction and operation of the LIGO Laboratory and Advanced LIGO as well as the Science and Technology Facilities Council (STFC) of the United Kingdom, and the Max-Planck-Society (MPS) for support of the construction of Advanced LIGO. Additional support for Advanced LIGO was provided by the Australian Research Council. We thank Dr. Andrew Matas and rest of the stochastic working group of LIGO for their help with the stochastic pipeline and their valuable comments. We acknowledge the use of IUCAA LDG cluster Sarathi for the computational/numerical work. This research benefited from a grant awarded to IUCAA by the  Navajbai Ratan Tata Trust (NRTT). A. A. acknowledges the support of Council of Scientific and Industrial Research (CSIR), India. S. M. acknowledges support from the Department of Science \& Technology (DST), India provided under the Swarna Jayanti Fellowships scheme.

\bibliography{manuscript}

\end{document}